\documentclass[aps,prd,10pt,
               preprintnumbers,numbers,sort&compress,nofootinbib,
               twocolumn,
               colorlinks, linkcolor=blue,
               citecolor=blue]{revtex4}
\usepackage{epsfig}

\begin{document}

\title{Quark Matter Induced Extensive Air Showers}
\author{Kyle Lawson}
\affiliation{Department of Physics and Astronomy, University of
  British Columbia, Vancouver, BC, V6T 1Z1, Canada}
\email[E-mail:~]{klawson@phas.ubc.ca}
\date{\today}

\begin{abstract}
If the dark matter of our galaxy is composed of nuggets of quarks 
or antiquarks in a colour superconducting phase there will be a 
small but non-zero flux of these objects through the Earth's atmosphere. 
A nugget of quark matter will deposit only a small 
fraction of its kinetic energy in the atmosphere and is likely to be 
undetectable. If however the impacting object is composed of 
antiquarks the energy deposited can be quite large. In this case nuclear 
annihilations within the nugget will trigger an 
extensive air shower the particle content of which is  similar to that 
produced by an ultrahigh energy cosmic ray. This paper gives a qualitative 
description of the basic properties of such a shower. Several 
distinctions from an air 
shower initiated by a single ultra high energy nucleus will be described 
allowing these events to be distinguished from the 
cosmic ray background. The subtlety of these features may mean that 
some fraction of the high energy cosmic ray spectrum may in fact 
be due to this type of dark matter interaction. 

The estimated flux of dark matter nuggets and the energy deposited 
in the atmosphere are such that the Pierre Auger Observatory may 
prove an ideal facility to place constraints  on the flux of 
heavy quark matter objects. This paper attempts to highlight the 
best techniques to search for a quark matter signature through an 
extensive air shower signal. 
\end{abstract}

\maketitle

\section{Introduction}
\label{sec:intro}
\subsection{quark matter as a dark matter candidate}
It has been suggested that the dark matter may be composed of 
macroscopically large, strongly interacting, composite objects comprised 
of the light quarks of the standard model in a non-baryonic 
phase such as strange quark matter 
\cite{Witten:1984rs} or a colour superconducting phase \cite{Zhitnitsky:2002qa}. 
In the latter case the composite objects may be bound 
states of either quarks or antiquarks 
which are stable over cosmological 
time scales. While strongly interacting these objects remain ``dark" 
due to their large mass to surface area ratio and the correspondingly 
low number density required to explain the 
observed dark matter mass density.  The total 
baryonic charge of the composite object is
the dominant uncertainty in this model as it depends on the poorly understood 
physics of nugget formation (which occurs 
at the QCD phase transition.) 
A combination of theoretical and 
observational constraints suggest that the mean baryonic charge  
must exceed $10^{20}$ \cite{Zhitnitsky:2002qa}while the upper 
bound is dependent on the formation model 
and is not well constrained. 

A brief qualitative overview of the structure of a quark 
nugget is given in appendix \ref{sec:structure}. Further, more precise, details 
of various phases of quark matter are available in the reference 
given there.

Previous works have studied the observational consequences of 
the presence of quark matter within the galaxy. No contradictions 
are found with existing observations, in fact the 
emission produced by these objects may help to explain 
several anomalies in the galactic spectrum such as the 
strong 511keV line \cite{Knodlseder:2005yq}, \cite{Zhitnitsky:2006tu}, 
\cite{Oaknin:2004mn}
the COMPTEL excess at 10 MeV \cite{Strong:2004de}, 
\cite{Lawson:2007kp}, the diffuse x-ray background 
\cite{Muno:2004bs}, \cite{Forbes:2006ba}
and the WMAP ``haze" \cite{Finkbeiner:2003im}, 
\cite{Forbes:2008uf}. Based on the simplest models of the 
dark matter distribution and nugget interaction with the 
interstellar medium a best fit to the galactic spectrum in this 
analysis is found to favor a baryonic charge for the nuggets of 
$B \sim 10^{25}$.

\subsection{high energy ``cosmic rays" from quark matter}
The cosmic ray spectrum is now observed to extend to energies  
above $10^{20}$eV \cite{Abraham:2010mj}. The incredibly small flux of cosmic rays at 
these energies requires a correspondingly large detector  
to obtain useful statistics for these events. The aim of this work 
is to highlight the possibility that these detectors can also impose 
significant constraints on massive composite dark matter candidates. 
Composite objects composed purely of matter will deposit only 
a fraction of their kinetic energy in the atmosphere. The small energy scales 
involved do not allow for substantial particle generation and 
make direct detection unlikely. However, in the case of a nugget 
composed of antimatter the dominant 
interactions between the atmosphere and antiquark matter will be 
strong force mediated matter-antimatter annihilations. The hadronic 
shower resulting from these annihilations will be dominated 
by light mesons and their decay products. The energy deposited by such an event 
will be considerably larger than the nugget's kinetic energy and the 
resulting shower should be readily observable. 
As in the case of  a single ultrahigh energy
proton or ion a quark nugget impacting the earth's atmosphere will 
be observable through the extensive air shower which develops 
around the primary particle.  However, in  the model considered here
the shower is
driven not by the kinetic energy of the primary but by the energy released in 
matter antimatter-annihilations. This makes these 
events fundamentally different than 
the previously considered cases of highly accelerated dust  
or strangelets \cite{Madsen:2004ef}. Existing models of cosmic rays 
require an accelerator capable of providing sufficient kinetic 
energy for the primary particle to trigger an extensive 
air shower, in the present case no such 
accelerator is required as the shower is driven by 
energy released in nuclear annihilations. This allows a large 
air shower to develop despite the fact that the primary particle 
has a relatively small (galactic scale) velocity. 

This paper gives an overview of the process by 
which a quark nugget deposits energy in the atmosphere 
and the properties of the resulting extensive air shower. 
As in the case of an air shower initiated by a single ultrahigh 
energy cosmic ray these quark matter induced
showers arise through a very large number of hadronic interactions 
which necessarily cascade down to similar final state products.
As such the particle content of the shower, as observed at the earth's 
surface will be quite similar to that of a conventional shower.  
A detailed description of  the resulting 
air shower would require large scale numerical  
simulations (similar to those conducted for proton or nuclei initiated showers) 
which are beyond the scope of this work. In an attempt to keep the 
physical picture as clear as possible the body of this work focuses on only the 
most essential features of the shower rather than microscopic details which may 
be strongly dependent on the precise structure of the strong interactions at large 
densities.  While a quark matter initiated shower is in many ways similar to 
a cosmic ray air shower there are also several critical differences in both 
the geometry and the timescales involved. The final section of this work 
highlights these differences and discusses potential techniques for 
the detection of quark matter induced air showers. 

\section{Total flux}
\label{sec:flux}
The exact distribution of dark matter in the galaxy remains uncertain. 
Recent simulations indicate the possibility of significant structure 
at subgalactic scales \cite{Springel:2005nw} 
which could significantly affect the flux of dark matter through the earth. 
In the interest of simplicity the following analysis assumes a local density 
consistent with a smooth density profile and a velocity set by virial equilibrium. 
Under these assumptions the 
dark matter density in the neighborhood of our solar system is 
$\rho_{DM} \approx 1.5 ~$GeV/cm$^3$. Assuming 
that the effective mass of quarks in a colour superconductor 
is comparable to that of hadronic quarks this 
mass density translates to a number density of nuggets 
approximately given by,
$n \sim  B^{-1} ~cm^{-3}$.
Where $B$ is the total baryon number of the nuggets. 
The number density can then be combined with the 
mean galactic velocity $v_g \sim 200km/s$  to obtain 
a flux of nuggets at the earth's surface. 
\begin{equation}
\label{flux}
\frac{dN}{dA~dt} = nv_g \approx  (10^{25} km^{-2}~yr^{-1})~B^{-1} 
\end{equation} 
Based on this order of magnitude estimation nuggets with a 
baryonic charge distribution 
near that favoured by fits to the galactic spectrum will produce a 
flux comparable to that of cosmic rays near the GZK limit \cite{Greisen:1966jv},
\cite{Zatsepin:1966jv}. 
It is precisely this flux range that the Pierre Auger 
Observatory \cite{Fick:2003qp} was designed 
to study and, consequently, it is also capable of  
constraining the presence of  heavy quark matter 
in the cosmic ray spectrum. One might also consider 
looking for a quark nugget signal at large underground 
detectors however, as discussed in appendix \ref{sec:underground_mus} 
the larger surface area presented by Auger allows it to 
impose much tighter constraints. 

\section{Energetics} 
\label{sec:energetics}
This section gives an overview of the energy considerations related 
to a quark nugget induced air shower without focussing on the details 
of how this energy is deposited in the atmosphere. While an antiquark 
nugget contains a large amount of antimatter very little of it 
actually annihilates as the nugget traverses the atmosphere. Instead the 
annihilation rate is limited by the rate at which the nugget sweeps up 
atmospheric matter which is dependent on the cross sectional area of the 
nugget and the atmospheric density. At the earth's surface the integrated 
mass of atmospheric molecules is on 
the order of $1kg/cm^2$ while the nugget radius is generally found to be
on the order of $10^{-5} cm$. For these values, if all the atmospheric 
molecules striking the nugget annihilate completely, 
the energy produced while crossing the atmosphere is, 
\begin{equation}
\label{eq:DEnergy}
\Delta E = 2X_{at} ~\pi R_n^2 = 10^{26}eV \left( \frac{R_n}{10^{-5}cm} \right)^2
\end{equation}
This represents the total energy production from annihilations. The 
majority of this energy is thermalized within the nugget and will not 
take a readily observable form. It will also be shown that only a 
fraction of all molecules incident on the nugget
actually annihilate. Thus, the expression (\ref{eq:DEnergy}) 
represents a maximum energy available to the shower with the actual 
value likely to be several orders of magnitude smaller. 

For comparison the kinetic energy transferred to the atmosphere can 
be estimated by assuming that all molecules in the atmosphere are 
accelerated from rest to the typical nugget velocity of $200km/s$. 
\begin{equation}
\label{eq:Dkin}
\Delta T = \frac{1}{2}X_{at} ~ \pi R_n^2 v_n^2 = 10^{17}eV \left( \frac{R_n}{10^{-5}cm} \right)^2
\end{equation}
This is many orders of magnitude below the energy produced 
by annihilations and represents only a minuscule fraction of 
the total energy involved. Kinetic energy transfer may accelerate 
a large number of atmospheric molecules but will be a purely elastic 
process producing neither new particles nor significant amounts 
of ionization. For this reason the following discussion 
will deal with only the shower produced by antimatter nuggets 
and the energy transferred by inelastic collisions will be 
ignored. 

\section{Shower Components} 
\label{sec:shower_comp}
As stated above the quark matter induced shower will primarily 
arise from  the annihilation of atomic nuclei within the nugget. The 
main product of these annihilations will be light mesons (the 
exact composition of these mesons depends on 
the form of quark matter realized in the 
nuggets \cite{Son:1999cm}.) Given the relatively low momenta 
at which they are produced these strongly interacting modes are 
unlikely to escape across the quark matter surface. Instead, through a 
complex series of interactions, they will loose energy to the 
lighter modes of the superconductor. This process results in a collection of 
excited electromagnetically bound modes as well as thermalizing 
energy within the nugget. The following sections give a brief overview 
of the particle content generated in these interactions.
\subsection{electromagnetic shower}
\label{subsec:E&M} 
There are three primary mechanisms which will result in the 
emission of energetic photons from the nugget. First 
annihilations within the nugget cascade from the initial mesons 
down to the leptonic modes. As the lightest available energy 
carriers the positrons within the quark matter absorb the majority of this 
momentum. A positron incident on the quark matter surface from 
within the nugget will rapidly decelerate within the strong electric 
fields at the surface and remain bound to the nugget. 
This process leads to the emission of x-rays 
through bremsstrahlung. A second radiation production 
mechanism involves energetic electrons produced inside the 
nugget which annihilate with the positrons of the 
electrosphere. These annihilations, as 
well as annihilations of the electrons of 
atmospheric molecules,  produce gamma rays with energies 
up to a few tens of MeV which will be released into the atmosphere. 
A final photon contribution comes from thermal emission from 
the surface of the electrosphere. As the nugget heats up due to 
the increasing rate of annihilations the 
surface can reach temperatures at the keV scale. This will result 
in the emission of considerable amounts of thermal radiation. 
These energetic photon components of the nugget emission 
spectrum will generate an electromagnetic shower as the ionize the 
surrounding atmospheric molecules.
\subsection{muons} 
\label{subsec:muons}
As mentioned above the electrons and positrons produced in the 
nugget are unlikely to be able to escape into the atmosphere. 
Muons, because of their larger mass, lose energy less efficiently 
and are able to escape from the nugget's surface. As such they 
are the dominant charged particles deposited in the atmosphere. 
Initial muon energies will be determined by the energy 
scale of the lightest hadronic 
modes of the colour superconductor, typically around a few hundred 
MeV. After escaping the nugget these muons lose energy to 
the surrounding atmosphere, generating fluorescence light in the 
process, until they decay into energetic electrons. The 
treatment of muon energy loss to the surrounding atmosphere is described 
in appendix \ref{sec:mu_prop} and is important in determining the 
morphology of the resulting shower.

The exact geometry of muon emission from the nugget 
is a complex problem. At a basic level the majority of atmospheric 
molecules first strike the nugget surface on the downward 
directed face. The molecules will have relatively little time to migrate 
across the surface before they penetrate into the quark matter and 
annihilate. As discussed in \cite{Forbes:2006ba} the combination 
of large penetration depth and the rapid energy loss from 
the jets produced by annihilations within 
the nugget favors the emission of muons directly perpendicular 
to the quark matter surface above the point of annihilation. 
This argument, when combined with the preferential flux of atmospheric 
material along the axis of the nugget's velocity implies preferential emission 
in the forward direction. The simplest model would imply something 
like a cosine dependence but an exact estimate of this effect would 
depend on quite complicated material transport properties near the 
surface. In what follows it will simply be assumed that emission 
preferentially occurs from the forward directed face of the nugget. 

\section{Nugget Thermodynamics}
\label{sec:therm}
Before proceeding to a more detailed description of a quark nugget 
induced air shower some basic thermodynamic properties of the 
nuggets must be introduced. The majority of the energy deposited 
by nuclear annihilations is thermalized within the 
nugget. The exact fraction, hereafter labeled $f_{T}$, is dependent 
on the exact details of the quark matter and will not be calculated here. 
As the annihilations happen at low momenta the products are likely to 
be emitted without a preferred direction and any energy moving 
deeper into the nugget will certainly be thermalized. This basic geometric 
consideration suggests that $1 < f_{T}< 1/2$ with values near the upper 
limit more likely.
 
 \subsection{thermodynamic equilibrium}
 \label{sec:therm_eq}
This thermal energy is eventually radiated from the nugget's 
surface at the point where the electrosphere becomes transparent 
to thermal photons. This process was described in \cite{Forbes:2008uf}  
where the emission spectrum was found to be 
\begin{equation}
\label{eq:thermal_emis}
\frac{dE}{dt~dA} \approx  \frac{16}{3} \frac{T^4 \alpha^{5/2}}{\pi} \sqrt[4]{\frac{T}{m_e}}
\end{equation}
implying a supression of thermal emission, with respect to blackbody, 
at low temperatures. The following analysis  assumes that 
thermalization happens rapidly enough that the nugget remains near 
thermodynamic equilibrium. Under this assumption the rate at 
which thermal energy is deposited by annihilations will be 
equal to the rate at which energy is radiated from the electrosphere. 
The accretion rate is set by the nugget's velocity and the local 
atmospheric density and allows the nugget's surface temperature 
to be determined at a given height. 
\begin{eqnarray}
\label{eq:Temp}
\left( \frac{T}{m_e} \right)^{17/4} &=& \frac{3\pi\alpha^{1/2}}{64} \frac{a_b^3}{m_e}
\rho_{at}(h)v_n f_{T} \nonumber \\
&=& \left(\frac{\rho_{at}(h)}{860g/cm^3}\right) \left( \frac{v_n}{200km/s}\right) f_{T}
\end{eqnarray}
This estimation should remain valid as long as the temperature remains 
well below the electron mass (which is true over the entire atmosphere.) 
This implies that the temperature of a nugget near the earth's 
surface will be around 20 keV provided that all material in the 
nugget's path is annihilated. 

\subsection{molecular deflection}
\label{sec:mol_def} 
This section is devoted to determining the maximum rate at which 
matter can be deposited onto a quark matter surface. Intuitively as 
the flux of matter onto the nugget's surface increases so must the rate 
at which the resulting energy is transfered away from the surface. 
While the exact mechanism by which this energy transfer occurs may 
be quite complicated any plausible outward transfer of energy will 
exert a pressure on the incoming matter and limit the rate at 
which it can be fed onto the quark surface. This negative 
feedback suggests that there will be a density beyond 
which the annihilation rate saturates.
The following analysis attempts to be as general as 
possible to extract a generic scale at which matter 
annihilation rates reach a maximum. 

As demonstrated in \cite{Forbes:2009wg} electron-positron 
annihilations at low temperature are dominated 
by the formation of an intermediate positronium state. 
Positronium formation is a resonance process with a probability 
near one at low momenta but which falls off rapidly as the 
centre of mass momentum of the collision is increased. If 
the momentum is substantially larger that the positronium 
binding energy ($2m_e\alpha$) then the probability of 
forming a positronium bound state becomes negligible. 
This happens very high in the atmosphere so that the primary 
annihilation channel at relevant atmospheric densities is 
the direct $e^+e^- \rightarrow 2\gamma$ process described 
in \cite{Lawson:2007kp}. At temperatures below the electron 
mass this process is actually less efficient than elastic 
scattering. In this case many positrons will scatter off of the 
incoming molecule before any of the electrons annihilate. 
The incoming molecules carry a kinetic energy 
$T_{at} = \frac{1}{2}M_{at}v^2$, for a nitrogen molecule  
striking the nugget at 200 km/s this energy is a few keV.  
As the temperature increases each positron scattering 
transfers more energy until the energy transfer 
becomes sufficient to deflect the incident molecule. 
The exact temperature at which this occurs is dependent 
on the exact  details of energy transfer within the electrosphere 
and will not be determined here. Instead the following analysis 
will simply assume that the temperature must be slightly 
above the kinetic energy of the incoming molecule. 

\section{Fluorescence profile}
\label{sec:fluo_pro}
This section attempts to map the thermodynamic evolution 
described above onto a physical description of the resulting air 
shower. The atmospheric fluorescence yield 
of a shower is determined primarily 
from the number of charged particles moving through the 
atmosphere at a given point. These particles lose energy to the 
surrounding atmosphere exciting nitrogen molecules 
which subsequently radiate in the UV band. 

The fraction of muons per annihilated nucleon 
which escape the nugget depends on the precise 
details of the quark matter surface and on the mass 
of the lightest mesons in the dense quark matter (the decay of 
these being the primary muon production channel.) 
In vacuum $p\bar{p}$ annihilations 
produce a large number of pions. The uncharged 
$\pi^0$s decay  to photons 
while the charged pions decay to muons. As such an annihilation in vacuum 
typically yields between four and six muons. This should be taken as 
the upper limit for total muon production per nucleon annihilated 
though only a fraction of these muons manage to 
escape the nugget. Thus, the rate of muon production 
per annihilated nucleon, $\chi_{\mu}$, has a maximum possible value 
of order one while the actual value may be substantially lower. 
The uncertainty in $\chi_{\mu}$ is sufficient that the magnitude 
of the fluorescence yield is only weakly constrained at the present level 
of analysis. 

\subsection{geometry}
\begin{figure}[t]
\begin{center}
\includegraphics[width = 0.4\textwidth]{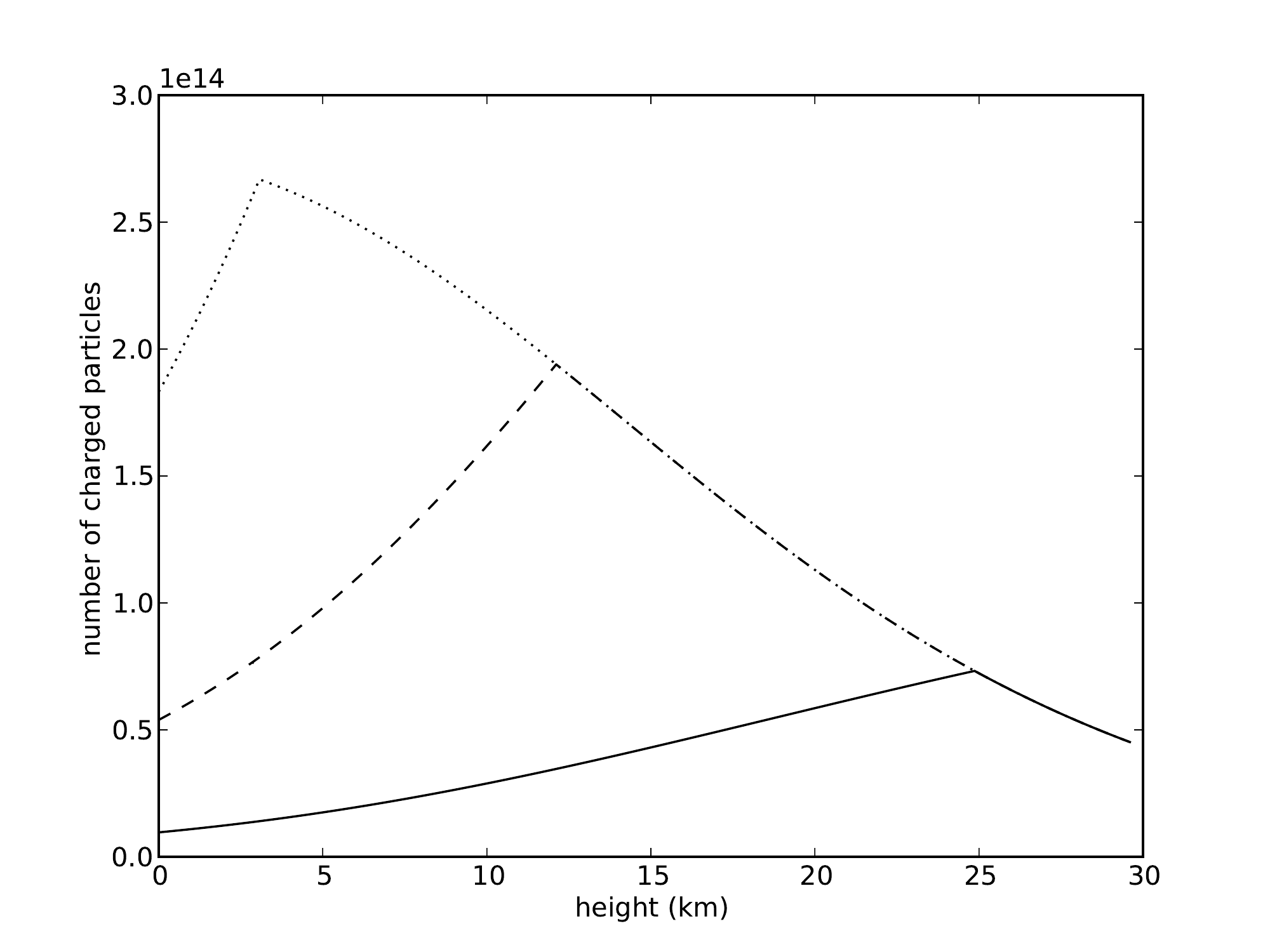}
\caption{Muon content of a quark matter initiated shower as 
a function of height. The curves are for saturation temperatures 
of 10keV (solid), 15keV (dashed) and 20keV (dotted).}
\label{fig:fluor_prof}
\end{center} 
\end{figure}

In section \ref{sec:mol_def} it was argued that there must be a 
temperature at which the nuclear annihilation rate saturates. If this 
happens at a nugget surface temperature $T_{max}$ then 
this rate may be found from expression \ref{eq:thermal_emis}. 
\begin{eqnarray}
\label{eq:rate_max}
\frac{dN}{dt} &=& \frac{32}{3} R_n^2\alpha^{5/2} \frac{T_{max}^4}{m_p}
\sqrt[4]{\frac{T_{max}}{m_e} } \nonumber \\
&\approx& 2\times10^{17}s^{-1} \left( \frac{R_n}{10^{-5}cm} \right)
\left(\frac{T_{max}}{10keV}\right)^{17/4}
\end{eqnarray}
Once this saturation point has been 
reached the decrease in the mean free path of an emitted particle 
with increasing atmospheric density implies that the flux 
of charged particles will decrease with atmospheric depth. 
The resulting shower profile, using the crude muon propagation model 
of \ref{sec:mu_prop} is shown in figure \ref{fig:fluor_prof}. 
It should be noted that the overall normalization of figure 
\ref{fig:fluor_prof} is highly uncertain as it depends on 
both the muon production rate $\chi_{\mu}$ and the mean 
energy with which muons escape the surface. Neither of 
these quantities are constrained beyond rough order of 
magnitude estimates.  Rather it is the 
overall geometry of figure \ref{fig:fluor_prof} that is relevant. 

The initial rise in muon flux is due to the increasing rate 
of nuclear annihilations with atmospheric density. The maximum 
charged particle number occurs near the point where the annihilation
rate saturates and, as the atmospheric density increases beyond 
this point, its main effect is to decrease the mean free path of a 
traveling muon. This results in a more rapid loss of muons from the 
shower and thus a decrease in the integrated charged particle flux. 

\subsection{timing}
This basic shower geometry,  growing to a maximum 
particle content then decreasing rapidly beyond that maximum, 
is similar to that associated with an ultrahigh energy cosmic ray shower, 
however the fluorescence timing will be substantially different. 
This difference arises due to the relatively small velocity of 
the nugget as compared to an ultra high energy cosmic ray. 
The later travels at the speed of light while the nuggets 
have typical galactic velocities, on the order of a few hundred 
kilometers per second, some three orders of magnitude slower. 

In both cases the secondary particles, produced in hadronic 
interactions, move outward at nearly the speed of light. As discussed in 
appendix \ref{sec:mu_prop} the charged particles of a quark matter induced 
shower are generally confined to a region within a few kilometers of 
the nugget due to their relatively small boost factors. 
The charged particles spread through this volume over the course 
of tens of microseconds. However, the illuminated region of 
atmospheric fluorescence will track with the nugget as it moves 
through the atmosphere with the shower front advancing quite 
slowly. The time scales for the progress of the nugget itself 
will be on the order of a tenth of a second. 

The long duration of the atmospheric fluorescence 
and the large photon multiplicity at any given time 
make these events very difficult to observe above 
the various backgrounds. For this reason 
the fluorescence detector of the Pierre Auger Observatory 
is unlikely to trigger on a quark nugget air shower 
\cite{Schmidt:2008uz}. The difficulties inherent in 
detecting these fluorescence events likely favors  
searches based on surface detectors. 
 
\section{Lateral surface profile} 
\begin{figure}[t]
\begin{center}
\includegraphics[width = 0.4\textwidth]{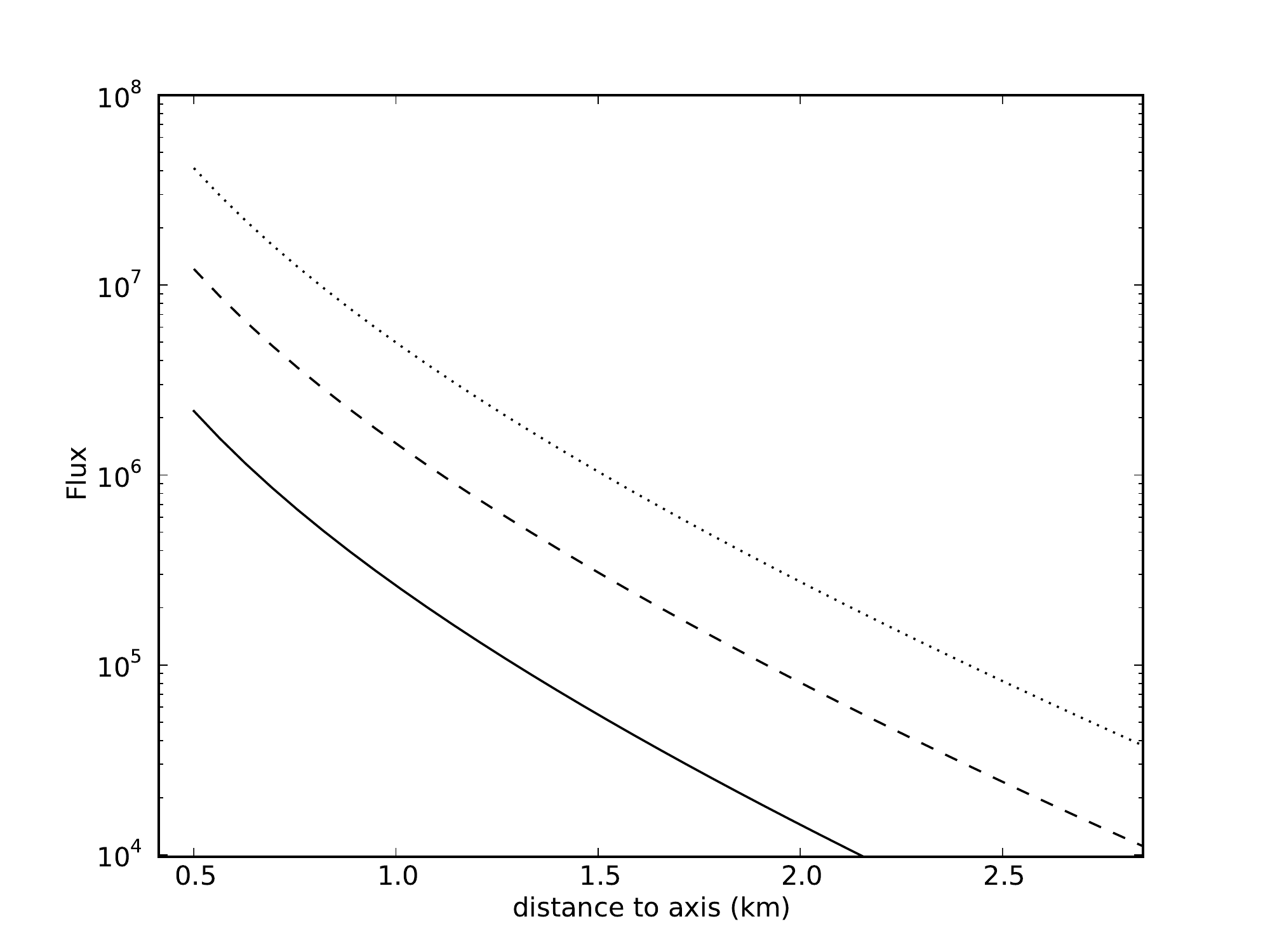}
\caption{Particle flux per $m^2$ as a function of distance from the 
shower core. The curves are for saturation temperatures of 
10keV (solid), 15keV (dashed) and 20keV (dotted).} 
\label{fig:lat_prof}
\end{center}
\end{figure}

When the shower reaches the earth's surface it will 
be tightly clustered around the nugget with only the highest 
energy shower components able to travel far from 
the shower core. As with the fluorescence profile the exact 
details of the lateral profile are dependent on models of 
muon propagation through the atmosphere. Again the 
results described here are based on the approximations 
of appendix \ref{sec:mu_prop} which intends only to capture the most 
general features of the shower. As the majority of muons are 
emitted at relatively low ($\sim 10MeV$) energies they are 
unable to travel far from the nugget in the dense 
lower atmosphere.  However, the shower also contains a 
smaller number of high energy muons able to travel 
a larger distance from the nugget. These higher energy 
muons produce an extended lateral distribution of particles 
at the surface. An approximate lateral profile of the shower 
is plotted in \ref{fig:lat_prof}. As with the fluorescence profile 
the total flux may be rescaled by slight changes in the 
muon production rate and spectrum. The scaling of figure 
\ref{fig:lat_prof} is therefore less significant than the general 
profile shape. 
The essential feature of the radial surface profile is a 
strong peak near the point where the nugget strikes the 
ground and an exponential drop off with radial distance from this 
point. The controlling scale for the exponential fall off is determined 
by the mean free path of a muon averaged over the allowed initial 
energy scales as described in \ref{sec:mu_prop}. Numerically 
it is found that this scale is in the range from a few hundred meters 
up to a few kilometers for the muon spectra given.
\subsection{timescales}
As with the fluorescence profile described above the surface 
particle distribution is similar in geometry to that of an air shower 
initiated by a single high energy cosmic ray. But, once again, 
the timing signature will be very different. In the case of a conventional 
air shower the particles (primarily muons) arrive at the surface within 
a timescale of less than a microsecond. This is particularly true of the 
strongly beamed particles quite near the shower core while the 
arrival times of particles far from the shower core show considerably 
more scatter. 

In the case of a quark nugget initiated shower the time scale 
for particle arrival is determined by how long it takes the nugget to 
pass through the region from which the emitted muons are able to 
reach the surface. As discussed above the critical length scale 
for muon propagation is on the order of several hundred meters. 
For a nugget moving at 200km/s this implies a shower duration 
on the order of several milliseconds, several orders of magnitude 
slower than the duration of an ultrahigh energy primary initiated 
shower. 

Near the shower core the difference in timing 
signatures between an ultrahigh energy 
cosmic ray shower and a quark nugget shower will be very clear. However, 
in the case of an off axis shower the situation is less clear. 
At larger radial distances the secondary particles of a cosmic ray shower 
are less strongly beamed and have undergone a larger number of 
scatterings resulting in a longer shower duration. The opposite is true in 
the case of a quark nugget induced air shower. In this case it is only the 
highest energy muons able to travel far from the shower core and 
the shower duration may be significantly shorter than near the shower core. 

\section{Comparison with conventional showers}
To this point emphasis has been placed on the similarities 
between the air shower induced by an antiquark nugget and 
one produced by a single ultra high energy primary. There are 
however several important distinguishing features between the 
two.  The most important of these arise from the much 
lower velocity of the primary particle.  

${\bullet}$ A longer shower duration will be observable in atmospheric fluorescence 
producing an extended fluorescence track which lasts for a longer time. 

${\bullet}$ This longer duration effect is likely also observable in 
the surface arrival times of secondary particles. Depending on the 
timing cuts on the surface detector data it is likely that the 
muons associated with the shower will continue to arrive (with decreasing 
frequency) over times on the order of microseconds.

${\bullet}$ The lower velocity of the primary particles will result in a 
correlation between the arrival direction and the 
direction of earth's motion with respect to the galaxy. This effect  
produces both seasonal variation (similar to that searched for 
in the DAMA experiment \cite{Belli:1999nz}) as well as a correlation 
with the direction of somotion around the galactic centre. 

${\bullet}$ The arrival direction of quark nuggets is determined 
by the local dark matter distribution and, as such, should show 
no correlation with galactic or nearby intergalactic objects. 
The presence of a quark matter component in the 
cosmic ray spectrum would thus dilute any existing correlation with 
the source of typical ultrahigh energy cosmic rays. 

${\bullet}$ A distinguishing feature unrelated to the primary particle's veloctiy
is that shower evolution is dependent on the surface temperature 
of the nugget. As may be seen in equation \ref{eq:Temp} this 
scales with the atmospheric density rather than the atmospheric 
depth of the shower. Conversely the evolution of a conventional shower 
is determined purely by the amount of atmospheric material through 
which the shower has propagated. A possible consequence of this 
effect would be a larger apparent depth of maximum for steeply 
inclined showers. However, without a detailed description of the 
thermal physics of the nuggets it is possible that the statistical 
variation in the saturation temperature may be large enough to 
obscure this effect. 

${\bullet}$ A final distinguishing feature is observable in muon spectroscopy. 
In both cases the majority of particles will be generated via the
decay of pions with QCD scale energies however an ultra high energy 
primary may produce a number of muons with energies well 
above this scale. Conversely the QCD scale sets the highest 
energy available to individual particles in a quark nugget 
initiated shower. An analysis of the muon spectrum at the 
surface will thus show a high energy cut off around a GeV
in the case of a quark nugget initiated shower while a 
conventional shower will show no such cut off.    

\section{Conclusion and Discussion}
The main purpose of this work has been to point out 
that large surface area cosmic ray detectors are also 
well suited to search for the presence of 
dark matter in the form of quark nuggets. The impact of 
an antiquark nugget on the atmosphere will produce an extensive 
air shower consisting of a large number of secondary 
particles observable through both their impact 
on surface detectors and the atmospheric fluorescence they 
generate. The resulting air shower is morphologically 
similar to one generated by a single ultra high energy primary 
particle in both the fluorescence profile and the lateral distribution 
at the earth's surface. It is therefore possible that some part 
of the high energy cosmic ray spectrum may arise from the partial
annihilation of dark matter in the form of heavy quark nuggets. 

The exact location of the shower maximum is dependent on 
rather complicated thermal physics in the electrosphere of the 
nuggets and, as such, cannot be explicitly formulated in the 
preliminary treatment presented here and will be the subject of 
future work.  From this analysis it is only possible to 
argue that there must be an atmospheric density at which 
thermal pressure overcomes the kinetic energy of atmospheric  
molecules causing the annihilation rate to saturate. This effect 
leads to a nontrivial height at which the shower will have a 
maximum particle content. In this context the observed break 
in the energy spectrum $10^{19.5}eV$ \cite{Abraham:2010mj} 
imposes limits on the total particle content and saturation 
temperature of the quark nugget. 

Finally it should be highlighted that additional work is needed 
on the atmospheric propagation of particles within this model. 
While the required simulations are simplified by the absence 
of very high energy interactions (the properties of which are 
not well established) the injection of particles is 
dramatically different from a conventional shower. 
This requires a fundamentally different formulation of the  
shower simulations from those presently employed. Without 
such simulations the extraction of statistical properties of the 
showers is not possible. 

\section{Acknowledgments}
Parts of this work were motivated by early discussions with Brian Fick, I
would also like to thank St\'{e}phane Coutu for many useful comments 
and Ariel Zhitnitsky for his helpful discussions.  
This research was supported in part by the Natural Sciences and Engineering 
Research Council of Canada.

\bibliographystyle{h-physrev}
\bibliography{cosmicrays}

\appendix
\section{Quark nugget structure}
\label{sec:structure}
As alluded to above there are several possible phases 
of quark matter from which the nuggets may be formed. Rather than 
performing detailed calculations within the context of a 
particular model this paper will rely only on general 
properties of quark matter. Reviews of these ideas are 
available  in several previous works such as 
\cite{Alcock:1986hz}, \cite{Madsen:2001fu}, \cite{Alford:2006fw}. 
This section will present only the minimal details necessary 
for the discussion of the phenomenology of quark nugget initiated 
air showers. 

The nuggets have a density at 
the nuclear scale and may 
have a lower binding energy than the iron nucleus. If this is the case 
nuggets formed in the early universe will be stable over 
cosmological time scales.

Of particular interest here is the proposal of \cite{Zhitnitsky:2002qa} 
in which the nuggets may be composed of both matter and antimatter. 
The preferential formation of anti-nuggets has been proposed as 
a mechanism for baryogenesis \cite{Oaknin:2003uv}. In this model the 
formation of anti-nuggets is favored by a factor of 3:2 so that, 
beginning from a universe with no net baryonic charge, antimatter 
is preferentially hidden in the dark matter nuggets \cite{Oaknin:2003uv}. 

At asymptotically large densities quark matter is 
composed of equal numbers of u,d and s quarks and is
charge neutral. However, the large s quark mass results 
in a depletion of s quarks in lower density quark matter. 
Even if the bulk of the nugget is charge neutral the 
decreasing density near the quark surface 
results in a depletion of s quarks and gives the quark matter 
a net charge, positive in the case of a matter nugget and 
negative in the case of an anti-nugget. To maintain 
charge neutrality the quark matter is surrounded by a 
layer of leptons. These leptons are only electromagnetically 
bound to the surface and extend beyond the quark surface. 
The exact structure of this layer, known as the 
electrosphere, was worked out in \cite{Forbes:2009wg}. 
Near the quark matter surface the electrons (or positrons) are tightly 
bound and at nuclear densities however the density falls off 
with distance down the atomic scale. The presence of a 
large atomic density shell of positrons surrounding the nugget 
will play a critical role in interaction between the nugget and 
molecules of the atmosphere. This layer also determines the 
thermal properties of the nugget as it is the point where the nugget 
first becomes transparent to low energy thermal photons.

\section{Muon propagation}
\label{sec:mu_prop}
This section gives a brief description of the approximations made in 
describing the evolution of the air shower. While the model used 
is very simple it is intended only for demonstrative purposes and 
highlights only the most basic properties of the shower. As described 
above, the only charged particles capable of escaping the quark nugget 
are muons. The main muon production channel is the decay of a 
meson-like excitation which will produce muons with energies at the 
GeV scale. These muons rapidly lose energy in subsequent scatterings,
primarily with the positrons which are the lightest available modes. 
Energy loss will continue until the momentum of the muon is on 
the same scale as the plasma frequency within the quark matter. 
This plasma frequency is generally found to be of the order
$\omega_p \sim e \Lambda_{QCD} \sim$ 
10 MeV for a wide range of quark matter phases \cite{Alcock:1986hz}.  
The muon energy spectrum will therefore be peaked at this energy 
but may run up to the GeV scale for muons directly produced in 
annihilations near the surface. Energy loss scales 
exponentially with the depth at which the muon is produced thus the 
energy spectrum will be approximated as, 
\begin{equation}
\label{eq:mu_spec}
\frac{dn_{\mu}}{dk} = \frac{1}{\omega_p}  e^{(\omega_p-k)/\omega_p}, ~~~
m_p > k > \omega_p
\end{equation}
where $k$ is the muon momentum and $m_p$ is the proton mass. 
This will be taken as the initial spectrum for muons escaping the 
nugget. 

A muon traveling through the atmosphere will lose energy  scattering 
off the surrounding molecules. As these are neutral on scales larger than 
a few times the Bohr radius scattering requires the exchange of photons 
with an energy above $m_e \alpha$. The cross-section for this processes 
in the limit where $m_{\mu}>>m_e\alpha$ is given by,
\begin{equation}
\label{eq:mu_xsec}
\sigma_{\mu,e} \approx \frac{2\pi \alpha}{m_e^2}\frac{1}{v^2} \equiv
\sigma_0 v^{-2}
\approx 7\times 10^{-23}cm^{2}\left( \frac{c}{v} \right)^2
\end{equation}
where v is the muon's velocity. This translates to a 
scattering length of 
\begin{equation}
\label{eq:l_scat}
l_s = \frac{1}{\sigma_{\mu ,e} n_{at}(h)}
\end{equation}
where $n_{at}$ is the number density of the atmosphere. 
Scattering losses are dominated by events involving the lowest 
possible intermediate energy photon, thus the muon will lose 
roughly $m_e\alpha$ worth of energy on scattering.
A muon with initial kinetic energy $T = E - m_{\mu}$ will 
then lose most of its energy after  $T/m_e\alpha$ scatterings. Thus 
the stopping length for a muon of energy E and momentum p will 
be 
\begin{equation}
\label{eq:L_stop(p)}
L_s = \frac{E-m_{\mu}}{m_e\alpha}l_s = 
\frac{E-m_{\mu}}{m_e\alpha}\left( \frac{p}{E} \right)^2 \frac{1}{\sigma_0 n_{at}}
\end{equation}
The other relevant length scale is the typical distance that a muon 
travels before it decays
\begin{equation}
\label{eq:L_decay}
L_d = v\gamma \tau_{\mu} = \frac{p \tau_{\mu}}{m_{\mu}}
\end{equation}
where $\tau_{\mu}=2.2 \times 10^{-6}s$ is the muon lifetime.  
Once the muon decays to an electron or positron it will rapidly 
be lost in the electromagnetic component of the shower which 
this analysis makes no attempt to trace the evolution of. 
A muon thus travels a distance 
\begin{eqnarray}
L(p) = &L_d& ~~~if L_d<L_s \\
&L_s& ~~~ if L_s <L_d \nonumber
\end{eqnarray}
before its energy is dissipated into the electromagnetic 
shower. Note that scattering is the dominant stopping process 
for low momentum particles at large atmospheric densities 
while decays dominate high in the atmosphere and for 
higher energy muons. The relevant quantity for what follows 
is actually the energy averaged length obtained by integrating 
over the muon spectrum (\ref{eq:mu_spec}).
\begin{equation}
\label{eq:L_stop}
\bar{L} = \int_{\omega_p}^{m_p} \frac{dp}{\omega_p} 
L(p) e^{(\omega_p - p)/\omega_p}
\end{equation}
Given the annihilation rate $\Gamma_{an}$ the total number of 
particles produced at a given height may be estimated as 
\begin{equation}
\label{eq:NofH}
N(h) = \frac{\Gamma_{an}}{v_n} \chi_{\mu}\bar{L}
\end{equation}
This expression will be used in the context of (\ref{sec:fluo_pro}) 
in order to track the evolution of the shower's particle content. 
Similar considerations can be used to approximate the 
particle content at the earth's surface. Under the assumption that 
particle emission from the nugget is primarily along the nugget's 
direction of motion the number 
of muons reaching an area of the surface will be 
\begin{equation}
\label{eq:surf_flux}
\frac{dN}{dA} = \int_0^{\infty} \frac{dh}{2\pi (h^2+b^2)} 
\frac{\Gamma_{an}}{v_n} \chi_{\mu} \mathcal{F}
\end{equation}
where $b$ is the distance from the shower core and 
$\mathcal{F}$ is the fraction of initial muons which 
are able to propagate far enough to reach the surface. 
Both loss mechanisms produce an exponential 
extinguishing of the initial muon number, the characteristic 
length scale for this process is given by the energy averaged 
length in \ref{eq:L_stop}. 
\begin{equation}
\mathcal{F} = exp \left[ -\frac{\sqrt{b^2+h^2}}{\bar{L}} \right]
\end{equation}
The integration of \ref{eq:surf_flux} with this expression 
for $\mathcal{F}$ is used to approximate 
the surface flux of the shower. 

\section{Underground detectors}
\label{sec:underground_mus}
This section briefly discusses the constraints imposed on 
quark nuggets based on underground detectors.
While the muonic shower can be quite extensive in the atmosphere
the higher density of rock strongly limits the range over which 
the muons can travel. This can be seen by replacing the atmospheric 
density in \ref{eq:l_scat} with the density of surface rock.  In this case 
the scattering length drops by a factor of at least a thousand and 
the muons are absorbed quite close to their production site. This is, 
of course, precisely the reason why such experiments conducted under 
a large mass of shielding rock. Thus, the ability to constrain the density 
of quark nuggets scales almost directly with detector area (or the effective  
cross section presented by the cavity in which the detector is 
located.) If we apply this to 
a relatively large detector such as Super-Kamiokande \cite{Fukuda:2002uc} 
the effective detector size is limited to, at most on the order of 100m$^2$. In this case, 
even near the upper limit of the allowed flux (1/km$^2$/yr)  one would 
expect an event rate of only $\sim$ 1/century. As such the detection 
probability remains small even for experiments with run times of almost 
a decade. 

It should also be noted that most underground experiments work very 
hard block the influence of muons produced outside of the fiducial volume 
of the detector. Given these cuts intended to remove the radioactive decay 
background only nuggets passing very close to the detector would be 
capable of generating a sufficiently high muon multiplicity to result in 
detection. 
\end{document}